\documentclass[preprintnumbers,amssymb,aps,preprint]{revtex4}

\usepackage{graphicx}% Include figure files
\usepackage{epsfig}

\def\lsim{\raise0.3ex\hbox{$\;<$\kern-0.75em\raise-1.1ex\hbox{$\sim\;$}}}
\def\gsim{\raise0.3ex\hbox{$\;>$\kern-0.75em\raise-1.1ex\hbox{$\sim\;$}}}

\begin{document}
%\preprint{}

\title{Dimuon CP Asymmetry in $B$ Decays and $Wjj$ Excess \\ in Two Higgs Doublet Models}
\author{Bhaskar Dutta$^{1}$, Shaaban Khalil$^{2,3}$, Yukihiro Mimura$^4$ and Qaisar Shafi$^5$}
\affiliation{
$^1$Department of Physics, Texas A\&M University, College Station, TX 77843-4242, USA \\
$^2$Center for Theoretical Physics at the
British University in Egypt, Sherouk City, Cairo 11837, Egypt.\\
$^3$Department of Mathematics, Ain Shams University, Faculty of
Science, Cairo, 11566, Egypt.\\
$^4$Department of Phyiscs, National Taiwan University, Taipei, Taiwan 10617, R.O.C. \\
$^5$Bartol Research Institute, Department of Physics and
Astronomy, University of Delaware, Newark, DE 19716, USA.}
\date{\today}

\begin{abstract}
We analyze the puzzle of the dimuon CP asymmetry in $B_s$ decays
in two Higgs doublet models. We show that the flavor changing neutral current (FCNC)
induced by the Higgs coupling in a type III two Higgs doublet model provides a solution to the
dimuon charge asymmetry puzzle by enhancing the 
absorbtive part of the mixing amplitude $\Gamma_{12}^s$.
We investigate different experimental constraints
and show that it is possible to enhance $\Gamma_{12}^s$ in order 
to explain the dimuon asymmetry observed by D0.
This enhancement requires large Higgs couplings to the first
and second generations of quarks which may also explain the recent
3.2 $\sigma$ $Wjj$ excess observed by CDF.

\end{abstract}
MIFPA-11-14

\maketitle
%%%%%%%%%%%%%%%%%%%%%%%%%%%%%%%%%%%%%%%%%%%%%%%%%%%%%%
%

\baselineskip 20pt

\section{Introduction}
%
%Recently,
The D0 collaboration has measured the like-sign dimuon
charge asymmetry in semileptonic $b$-hadron decays $A^b_{sl}$. The
following result has been reported \cite{Abazov:2010hv}:%
\begin{equation}
A^b_{sl} = -0.00957\pm 0.00251 ({\rm stat})
\pm 0.00146 ({\rm syst}).%
\label{d0result}
\end{equation}
The like-sign dimuon charge asymmetry $A_{sl}^b$ for semileptonic
decay of $b$ hadrons is defined as%
\begin{equation}
A^b_{sl} = \frac{N_b^{++} - N_b^{--}}{N_b^{++} + N_b^{--}}, %
\end{equation}
where $N_b^{++}$ and $N_b^{--}$ are the number of events containing
two $b$-hadrons that decay semileptonically two positive or
negative muons with the quark subprocesses: $b\to \mu^- \bar\nu X$ and
$\bar{b}\to \mu^+ \nu X$.

This result indicates a 3.2 $\sigma$ deviation from the Standard Model (SM)
prediction. A confirmation of this deviation
would provide unambiguous evidence for new physics (NP) at
low energy with a new source of CP violating phases. It is a common
feature for any physics beyond the SM to possess additional
sources of CP violation besides the SM phase in quark mixing
matrix. These new phases can induce sizable contributions to
direct and indirect CP asymmetries in $B_{d,s}$ decays and thereby resolve
the apparent discrepancies between the observed results and the SM
expectations.

The charge asymmetry $A^b_{sl}$ at the Tevatron can be expressed as %
%\be %
$A^b_{sl} = (0.506 \pm 0.043) a_{sl}^d + (0.494 \pm 0.043) a_{sl}^s$
\cite{Abazov:2010hv},
%\ee%
where the charge asymmetry $a^q_{sl}$ ($q=d,s$) for ``wrong-charge"
semileptonic $B_q^0$-meson decay induced by the oscillation is
defined by
\begin{equation}
a_{sl}^q = \frac{\Gamma(\bar{B}^0_q \to \mu^+ \nu X) -
\Gamma(B^0_q \to \mu^- \bar\nu X)}{\Gamma(\bar{B}^0_q \to \mu^+ \nu X)
+ \Gamma(B^0_q \to \mu^- \bar\nu X)}. %
\end{equation}
This asymmetry can be written as \cite{Hagelin:1981zk}
\begin{equation}
a_{sl}^q = \frac{\vert \Gamma^q_{12} \vert}{\vert M^q_{12}
\vert} \sin \phi_q =\frac{\Delta \Gamma_q}{\Delta M_q} \tan \phi_q , %
\label{a_sl^q}
\end{equation}
where the mass and width differences between $B_q$ mass eigenstates are given by
\begin{eqnarray}
\Delta M_{B_q}&=& M_{B_H} - M_{B_L}= 2 \vert M_{12}^q \vert ,\\
\Delta \Gamma_{q} &=& 2 \vert \Gamma^{q}_{12} \vert \cos \phi_{q} ,%
\end{eqnarray}
and $M^q_{12}$ and $\Gamma^q_{12}$ are the dispersive
and the absorptive parts of the mixing amplitudes, respectively.
The CP violating phase $\phi_{q}$ is defined by
$\phi_{q}=\arg\left(-M_{12}^{q}/\Gamma^{q}_{12}\right)$.
In the SM,
the arguments of $M^q_{12}$ and $\Gamma^q_{12}$ are aligned at the leading order
due to the unitarity of the quark mixing matrix,
and thus, the phase $\phi_q$ is small
irrespective of the individual phases,
$\beta_d \equiv {\rm arg}(-V_{cb}^* V_{cd}/V_{tb}^* V_{td})$
and
$\beta_s \equiv {\rm arg}(-V_{tb}^* V_{ts}/V_{cb}^* V_{cs})$.
%
%$\beta_q$. %\equiv {\rm arg}(-V_{tb}^* V_{tq}/V_{cb}^* V_{cq})$.
The ratio of $\Gamma^q_{12}/M^q_{12}$ is
roughly proportional to $m_b^2/m_t^2$ with a small phase,
and as a result, the SM prediction of the charge asymmetry is small:
$a_{sl}^d(\rm{SM}) \simeq -5 \times 10^{-4}$,
$a_{sl}^s(\rm{SM}) \simeq 2 \times 10^{-5}$
and $A_{sl}^b(\rm{SM}) \simeq -0.00023$ \cite{Lenz:2006hd},
%
%$\phi_s \equiv \beta_s = \arg(V_{cs} V_{cb}^*/V_{ts} V_{tb}^*)
%\sim 0.02$ which leads to $a^s_{sl}(SM) \simeq  2.1 \times 10^{-5}$.
%
%Therefore, the SM prediction for $A_{sl}^b$ is given by
%$A_{sl}^b(SM) \simeq a^d_{sl}(SM)/2 \simeq -0.00023$,
which is in
clear contradiction with the D0 result in
eq.(\ref{d0result}).

In this respect, it is clear that a large new
CP violation in $B_s$ mixing is required to enhance $a_{sl}^s$
if $B_d$ mixing does not contain a new physics contribution.
The phase $\phi_s$ can be large in general in many new physics models
because the phase alignment between $M_{12}^s$ and $\Gamma_{12}^s$
can be broken
if a new particle, such as a supersymmetric particle, propagates
in the loop diagram, which contributes to the mixing amplitude \cite{Dutta:2010ji}.
%
%since the dispersive part of the mixing amplitude $M_{12}^s$
%can be easily modified in many new physics models
%
However, the absorptive part $\Gamma_{12}^s$ is not necessarily modified
by the propagation of new particles
due to the on-shell condition of the intermediate states.
%
%However,
The magnitude of $M_{12}^s$ is determined by the mass difference $\Delta M_{B_s}$,
and thus eq.(\ref{a_sl^q}) tells us that
$a_{sl}^s$ has a maximal value
if $\Gamma_{12}^s$ is dominated
by the SM tree level contribution \cite{Lenz:2006hd}.
One can easily show that
if one simply extrapolates eq.(\ref{a_sl^q}),
%if $\Gamma_{12}^s$ is dominated
%by the SM tree level contribution \cite{Lenz:2006hd},
then one needs a new phase
$\phi_s$ with $\sin \phi_s \gsim 1.6$ (which is outside of the
domain of $\sin\phi_s$) in order to account for
the experimental result of $A^b_{sl}$ \cite{Hou:2007ps} by the D0 collaboration.
Also with a large $\sin\phi_s \simeq 1$,
the decay width difference $\Delta \Gamma_s = 2 |\Gamma_{12}^s|\cos\phi_s$
is suppressed and
becomes inconsistent with the experimental constraints.
Therefore, one
concludes that the sustainability of the D0 results
of the like-sign
dimuon charge asymmetry in semileptonic $b$-hadrons decay would be
a clear hint of possible NP that modify the absorptive part
of the mixing amplitude %decay width
$\Gamma_{12}^s$.

 In this paper we show that the flavor changing neutral current (FCNC)
induced by the Higgs coupling
in two Higgs doublet models (THDM)  can provide a solution to the
dimuon charge asymmetry puzzle by enhancing %the decay width
$\Gamma_{12}^s$.
%in addition to generating a large new CP
%violating phase in $B_s$-$\bar{B}_s$ mixing.
The THDM is classified by the selection of the Higgs couplings to fermions.
We will consider a general type of coupling (so called type III THDM)
to obtain an appropriate FCNC source to modify $\Gamma_{12}^s$.
We enumerate the experimental constraints
and investigate if there is room to enhance $\Gamma_{12}^s$
to achieve the D0 result of the dimuon asymmetry.
We will obtain the operators
generated by charged Higgs exchange
which can enhance $\Gamma_{12}$.
To accomplish this, suitably large Higgs couplings to the first
and second quark generations needed.
Such couplings can, in addition, lead to an excess of
Higgs decays into dijets.
Indeed, the CDF collaboration
has recently reported a $3.2\sigma$ excess
in the 120-160 GeV range
in the invariant mass distribution
of the dijets in association with a $W$ boson \cite{Aaltonen:2011mk}.
The excess may be explained by the Higgs decays
to dijets via the new Higgs couplings.

The presence of two Higgs doublets is required
in supersymmetric models and in models with the left-right gauge symmetry
$SU(2)_L \times SU(2)_R \times U(1)_{B-L}$ \cite{moha}.
We briefly explore if the left-right
symmetric model can explain the large dimuon asymmetry consistent with other experimental
constraints.

\section{Dimuon asymmetry in two Higgs doublet models}

The two Higgs Doublet Model is  an extension of the SM that naturally
introduces a new source of CP violation and FCNC \cite{Buras:2010zm}.
In this
class of models, the most general renormalizable and gauge
invariant Yukawa interactions are given by%
\begin{eqnarray}
-{\cal L}_Y &=&
Y_u^{ij} q_i u^c_j H_u + Y_u^{\prime ij} q_i u^c_j \tilde H_d
+ Y_d^{ij} q_i d^c_j H_d \nonumber\\
&+& Y_d^{\prime ij} q_i d^c_j \tilde H_u
+ Y_e^{ij} \ell_i e^c_j H_d +
Y_e^{\prime ij} \ell_i e^c_j \tilde H_u,
\end{eqnarray}
where the Higgs fields $H_{u,d}$ have hypercharges $Y=\pm 1/2$, and
$\tilde H_{u,d}$ are defined as $\tilde H_{u,d}= -i \tau_2 H_{u,d}^*$. The
Yukawa couplings $Y_f$ and $Y'_f$ ($f=u,d,e$) are $3 \times 3$ matrices with a
generic flavor structure. In this case, the fermion masses are
given as
\begin{eqnarray}
M_u = Y_u^{ij} v_u + Y_u^{\prime ij} v_d, \\
M_d = Y_d^{ij} v_d + Y_d^{\prime ij} v_u, \\
M_e = Y_e^{ij} v_d + Y_e^{\prime ij} v_u.
\end{eqnarray}
In the basis where the mass matrix is diagonal, FCNC interactions through the
neutral Higgs ($H$ and $A$) exchange are generated from the mismatch
between the diagonalization of the mass matrices and the Yukawa
interactions. For instance, couplings like $s_L b_R^c H_d^0$ and
$b_L s_R^c H_d^0$ are induced by
$(V^d_L Y_d V_R^{d^+})^{ij} q_i d^c_j H_d
= (M_d^{\rm diag}/v_d - V^d_L Y_d^\prime V_R^{d^+} \tan\beta)^{ij} q_i d^c_j H_d$,
where $\tan\beta = v_u/v_d$, and
$V_{L,R}^d$ are the diagonalizing matrices of
$M_d$.
Hereafter, we will work in the basis where $M_d$ is diagonal,
and we omit $V_L^d,V_R^d$ in the expressions.
Then, by definition, $Y_d^{ij} = - Y_d^{\prime ij}\tan\beta$ ($i\neq j$).

These couplings contribute to $B_s$-$\bar B_s$ mixing
and modify the amplitude $M_{12}^s$ as follows \cite{Dobrescu:2010rh}:%
\begin{equation}
M_{12}^s = (M_{12}^s)^{\mathrm{SM}} + (M^s_{12})^{\mathrm{2HD}},
\end{equation}
where $(M^s_{12})^{\mathrm{2HD}}$ is given by%
\begin{equation}
(M^s_{12})^{\mathrm{2HD}}\simeq
\frac{Y_d^{23} Y_d^{32^*}}{m_H^2}
\langle B^0_s \vert (\overline{b_L} s_R)(\overline{b_R} s_L) \vert \bar{B}^0_s \rangle. %
\end{equation}
%
%Here we assume that $(V^d_L Y^{ij}_d V_R^{d^+})_{32} \simeq
%(V^d_L)_{33} Y_d^{32} (V^d_R)_{22} \simeq Y^d_{32}$.
%
We note that even in the minimal supersymmetric standard model (MSSM)
this type of modification of $M_{12}^s$ via the neutral Higgs exchange can be obtained
for large $\tan\beta$
through the finite correction of the Yukawa couplings
due to soft SUSY breaking terms.
In this case, if $\Gamma_s^{12}= (\Gamma_s^{12})^{\rm SM}$,
 the charge asymmetry $a^s_{sl}$ is given by%
\begin{equation}
a_{sl}^s = \frac{\vert (\Gamma_s^{12})^{\rm SM} \vert}{r_s \vert
(M_q^{12})^{\rm SM}
\vert} \sin (\phi_s^{\rm SM} + 2 \theta_s) , %
\end{equation}
where $r_s = \vert 1+ M_{12}^{\rm 2HD}/M_{12}^{\rm SM}\vert$ and
$2 \theta_s = \arg(1+M_{12}^{\rm 2HD}/M_{12}^{\rm SM})$.
Thus, $2 \theta_s$ can be large
if $\vert Y_d^{23} Y_d^{32*} \vert$ and
$\arg(Y_d^{23} Y_d^{32^*})$ are large,
satisfying the experimental constraint : $r_s \sim 1$.
As mentioned above, a large
value of $2\theta_s$ is necessary but not sufficient to account
for the dimuon CP asymmetry $A^b_{sl}$. A significant
enhancement of $\Gamma_{12}^s$ is preferable.

%As mentioned, we will study if $\Gamma_{12}^s$ can be modified
%in 2HDM.
%
In the 2HDM several $\Delta b=1$ effective operators which may
modify $\Gamma_{12}^s$, can be generated. For example, with
non-vanishing $Y_u^{\prime 32}$ and $Y_u^{\prime 22}$ the operator
%$b_L \bar s_L c_R \bar c_R$
$(\overline{\mathstrut b_L} c_R)(\overline{\mathstrut c_R} s_L)$
%$(\overline{b_L} c_R)(\overline{c_R} s_L)$
is generated through the charged
Higgs exchange.
Let us enumerate the possible $\Delta b = 1$ effective operators
 generated via the Higgs exchange, which are suitable for modifying $\Gamma_{12}^s$.
\begin{enumerate}
 \item $(\overline{\mathstrut b_L} u^i_R)(\overline{\mathstrut u^j_R} s_L)
= -\frac12 (\overline{\mathstrut b_L} \gamma_\mu s_L)
(\overline{\mathstrut u^j_R} \gamma^\mu u^i_R)$

 \item $(\overline{\mathstrut b_R} u^i_L)(\overline{\mathstrut u^j_L} s_R)
= -\frac12 (\overline{\mathstrut b_R} \gamma_\mu s_R)
(\overline{\mathstrut u^j_L} \gamma^\mu u^i_L)$

 \item $(\overline{\mathstrut b_R} s_L)(\overline{\mathstrut u^i_R} u^j_L)$
and $(\overline{\mathstrut s_R} b_L)(\overline{\mathstrut u^i_R} u^j_L)$

 \item $(\overline{\mathstrut b_R} s_L)(\overline{\mathstrut \tau_L} \tau_R)$
and $(\overline{\mathstrut s_R} b_L)(\overline{\mathstrut \tau_L} \tau_R)$

\end{enumerate}

The operator $(\overline{\mathstrut b_L} u^i_R)(\overline{\mathstrut u^j_R} s_L)$
is generated through the charged Higgs exchange with the coefficient:
\begin{eqnarray}
\sum_{q,q^\prime=u,c,t} \frac{V_{qb} V_{q^\prime s}^*}{m_{H^+}^2 \sin^2\beta}
\left[\frac{M_u^{\rm diag} \cos\beta}{v} - Y_u^\prime \right]_{qi}
\left[\frac{M_u^{\rm diag} \cos\beta}{v} - Y_u^\prime \right]_{q^\prime j}^*.
\label{bscc-operator}
\end{eqnarray}
As we will see below, this is the preferred operator for modifying $\Gamma_{12}^s$.

The operator $(\overline{\mathstrut b_R} u^i_L)(\overline{\mathstrut u^j_L} s_R)$
is generated through the charged Higgs exchange with the coefficient:
\begin{eqnarray}% \
\sum_{q,q^\prime=d,s,b}
\frac{V_{iq} V_{jq^\prime}^*}{m_{H^+}^2 \cos^2\beta}
\!\left[\!\frac{M_d^{\rm diag}\sin\beta}{v}\!-\! Y_d^\prime \right]_{q3}
\!\left[\frac{M_d^{\rm diag}\sin\beta}{v}\! -\! Y_d^\prime \right]_{q^\prime 2}^*\!.
\end{eqnarray}
This operator can modify $\Gamma_{12}^s$
through interference with $W$ boson exchange.
The effect can be large if $Y_d^{\prime 23}$ and $(Y_d \sin\beta-
Y_d^\prime \cos\beta)_{22}$ are sizable.
However, one needs fine-tuning to obtain a large
contribution to $\Gamma_{12}^s$
since the strange mass is $m_s = (Y_d \cos\beta+ Y_d^\prime
\sin\beta)_{22} v$.
Such fine tuning may also generate the
operator $(\overline{\mathstrut b_R} s_L)(\overline{\mathstrut s_L} s_R)$,
%$s_L \bar b_R s_L \bar s_R$
which affects $B_d \to \phi K$ by an amount which causes
disagreement with the experimental data.

The operators
$(\overline{\mathstrut b_R} s_L)(\overline{\mathstrut u^i_R} u^j_L)$
and $(\overline{\mathstrut s_R} b_L)(\overline{\mathstrut u^i_R} u^j_L)$
can be generated through the neutral Higgs exchange.
We note that both operators are needed to modify $\Gamma_{12}^s$
since $(\overline{\mathstrut u^i_R} u^j_L)$ cannot be self-conjugate.
%

% with the coefficient:
%
%Also, by the neutral Higgs exchange, $b_L \bar s_R c_L \bar c_R$
%and $s_L \bar b_R c_L \bar c_R$ operators can obtained. The $b_R
%\bar s_R c_L \bar c_L$ operator can be generated from $Y_d$ and
%$Y_d^{\prime}$ via the charged Higgs exchange. This operator can
%interfere with $W$ exchange and leads to modification for
%$\Gamma_{12}^s$. It gives the following contribution:

%

Finally, the operators $(\overline{\mathstrut b}_R s_L)
(\overline{\mathstrut \tau}_L \tau_R)$ and $(\overline{\mathstrut
s}_R b_L) (\overline{\mathstrut \tau}_L \tau_R)$ can be generated
from $Y_d^{23}$, $Y_d^{32}$ via the neutral Higgs exchange. It is
remarkable that when $Y_e^{\prime 22}v_u$ generates the muon mass,
Br($B_s \to \tau\tau$) is not constrained by experiment and
$\Gamma_{12}$ can be modified.  Note that in MSSM (or THDM with
$Y_e^\prime \to 0$), the following relation is obtained
\begin{equation}
\frac{{\rm Br}(B_s \to \tau\tau)}{{\rm Br}(B_s \to\mu\mu)} \simeq
\left(\frac{m_\tau}{m_\mu}\right)^2 .
\end{equation}
Thus, due to the experimental bound on Br($B_s\to\mu\mu$),
the quantity Br($B_s\to\tau\tau$) cannot be large enough to modify
$\Gamma_{12}^s$.

In the case of neutral Higgs exchanges,
$\Gamma_{12}$ is modified at one-loop level
while
$\Delta b=2$ operator to modify $M_{12}^s$ is also generated
through the neutral Higgs exchanges at tree level.
Therefore, to obtain a sizable $\Gamma_{12}^s$ contribution,
a large $M_{12}^s$ is also generated, which is unwanted.

%It is worth mentioning that the charged Higgs contribution from
%$b_L \bar s_L c_R \bar c_R$ operator does not necessarily modify
%$M_{12}^s$, but can modify $\Gamma_{12}^s$. Other contributions
%using the neutral Higgs exchange modify both $M_{12}^s$ and
%$\Gamma_{12}^s$. Since $\Delta M_s$ is consistent with the SM
%prediction, the charged Higgs contribution from $Y_u^\prime$ can
%provide an interesting solution. Actually, it should be if CP
%phase in $B_s \to J/\psi \phi$ decay is small (as indicated by CDF
%data from 2.8 fb$^{-1}$ to 5.2 fb$^{-1}$)

%Now let us study the experimental constraints.
Next we consider the following experimental constraints:
\begin{enumerate}
	\item $B\to M_1 M_2$ and $B\to \ell\ell$ decays.
	\item $b\to s\gamma$. %constraint
	\item lifetime ratio $\tau_{B_s}/\tau_{B_d}$.

\end{enumerate}

The constraints from two body decays into mesons and leptons are
studied in \cite{Bauer:2010dga},
and the allowed operators for $\Delta b=1$ are
\begin{equation}
\bar s b \bar c c, \quad \bar s b \bar \tau \tau, \quad \bar d b
\bar c u.\label{chargedhiggs}
\end{equation}
In the MSSM, it is difficult to enhance the $\Delta b = 1$ operators
selectively
because the interactions are related by known (or constrained) coupling constants.

The non-holomorphic %wrong term
Yukawa coupling $Y_u^{\prime}$ can also contribute
to the charged Higgs interaction and lead to an important effect
on $b\to s\gamma$. In a non-SUSY type II 2HDM, the charged Higgs has
to be heavier than $\sim 350$
GeV to agree with the experimental measurement of $b\to s\gamma$.
However, in our case, this constraint can be relaxed by a small
$Y_u^{\prime}$ contribution.
Note that $Y_u^{\prime 23}$ ($Y_d^{\prime 32}$)
is constrained by $b_R \to s_L \gamma$ ($b_L \to s_R \gamma$)
contribution.

%Due to the constraints from the decay modes: $B \to M_1 M_2$ and
%$B \to \ell \ell$, the allowed operators to give a sizable
%contributions to $\Gamma_{12}^{s,d}$ are \cite{xx}
%\begin{equation}
%\bar s b \bar c c, \quad \bar s b \bar \tau \tau, \quad \bar d b
%\bar c u.
%\end{equation}

The $b\to s\gamma$ process also constrains the large log contribution
due to the renormalization group evolution below the $W$ boson mass \cite{Borzumati:1999qt}.
The $b\to s\gamma$ constraint disfavors the neutral Higgs
contributions.
This is due to the fact that the operators
$(\bar s_L b_R)(\bar f_L f_R)$ and $(\bar s_R b_L)(\bar f_R f_L)$ $(f=c,b,\tau)$
give leading order large log corrections ($\delta C_{7,8} \sim m_f/m_b \ln M_W^2/m_b^2$)
to the $b\to s\gamma$ operators, and hence they are dangerous operators.
We note that the charged Higgs contribution can generate only next to leading
order corrections.

Finally, the lifetime ratio $\tau_{B_s}/\tau_{B_d}$ provides a
stringent constraint on the operator contributing to
$\Gamma_{12}^s$. The modification of $\Gamma_{12}^s$ can induce a
change ($\sim O(10)$\%) in the lifetime of $B_s$, which can be
consistent with the large hadronic uncertainty.
However, this uncertainty is cancelled in the lifetime ratio
$\tau_{B_s}/\tau_{B_d}$.
The current world average of the experimental result is \cite{Aaltonen:2011qs}
\begin{equation}
%\tau_{B_s}/\tau_{B_d} = 0.965 \pm 0.017.
\tau_{B_s}/\tau_{B_d} = 0.99 \pm 0.03.
\label{lifetime-ratio}
\end{equation}
Therefore, if $\Gamma_{12}^s$ is modified,
the lifetime of $B_d$ should also modified,
%
%$\Gamma_{11}^d$ should
%be also modified.
%
which provides a strong constraint.
Actually, for $B_d$ system, it is known that the bound for
 Br($B_d \to \tau \tau) < 4.1 \times 10^{-3}$.
Consequently, the modification of $\Gamma_{12}^s$
via $B_s \to \tau\tau$ has a deficit due to this lifetime ratio constraint.
The only allowed operator to modify the $B_d$ lifetime seems to be $\bar d b \bar c u$,
which can modify $\Gamma_{11}^d$.
Fortunately, the operator $\bar d b \bar c u$ alone
modifies neither $M_{12}^d$ nor $\Gamma_{12}^d$
since $\bar c u$ is not self-conjugate.

The above constraints imply that the only possibility left is the
charged Higgs exchange operator
$(\overline{\mathstrut q_L}\gamma_\mu b_L)
(\overline{\mathstrut u^i_R} \gamma^\mu u^j_R)$ $(q = d,s)$.
The coefficients of these operators are obtained,
as in eq.(\ref{bscc-operator}),
and they are proportional to $X_{ib}X_{jq}^*$,
where $X_{ij} = [(M_u^{\rm diag}\cos\beta/v - Y_u^\prime)^{\rm T} V]_{ij}$.
In order to generate the operators
$\bar s b \bar c c$ and $\bar d b \bar c u$,
we need $X_{cs}, X_{cb}$ and $X_{ud}$.
The condition $X_{cs} \sim X_{ud}$ can make these
effective operators comparable to keep the lifetime ratio $\tau_{B_s}/\tau_{B_d}$.
The condition can be satisfied when $Y_u^{\prime 11} \sim Y_u^{\prime 22}$.
However,
if $Y_u^{\prime 11}$ is sizable, one needs fine tuning to obtain the proper up quark mass.
This can be relaxed if $\tan\beta$ is large.
%

%From the leading order estimation, one finds
%\begin{equation}
%\frac{M_{12}^{\rm NP}}{M_{12}^{\rm SM}} \sim \left(\frac{X_{cs}^*
%X_{cb}}{g^2 V_{ts}^* V_{tb}} \right)^2 \frac{M_W^2}{M_H^2}
%\frac{f(m_c^2/M_H^2)}{S(x_t)},
%\end{equation}
%where $X_{ij} = [(M_u \cos\beta/v - Y_u^\prime)^{\rm T} V]_{ij}$
%and the loop function $f(x)$ is given by $f(x) = f(x,x,1)$, and
%$S(x)$ is the Inami-Lim function. The condition $X_{cs} \sim
%X_{ud}$ can make $\bar s b \bar c c$ and $\bar d b \bar c u$
%operators comparable to keep lifetime ratio. The condition $X_{cs}
%\sim X_{ud}$ is satisfied when $Y_u^\prime{}_{11} \sim
%Y_u^\prime{}_{22}$. However, because of the up quark mass, one
%needs fine-tuning to obtain sizable contributions of
%$\Gamma_{12}^s$. To relax the fine-tuning, larger $\tan\beta$ (or
%uplifted SUSY) is preferable. When one make $Y_u^\prime$ smaller
%to relax the fine-tuning, $m_H$ has to be lighter. Then,
%constraints from $B^- \to \tau \bar\nu$ and $B \to D \tau \bar\nu$
%will be severe.

The $\Gamma_{12}^s$ contribution is estimated as
\begin{equation}
\frac{\Gamma_{12}^{s \rm THDM}}{\Gamma_{12}^{s \rm SM}} \sim
\left(\frac{X_{cs}^* X_{cb}}{g^2 V_{ts}^* V_{tb}} \right)^2
\frac{M_W^4}{M_H^4} \gamma_{cc},
\end{equation}
where $\gamma_{cc} = \sqrt{1-4m_c^2/m_b^2}(1-2/3 (m_c^2/m_b^2))$.
The contribution to the dispersive part of the mixing can be written as
\begin{equation}
\frac{M_{12}^{s \rm THDM}}{M_{12}^{s\rm SM}} \sim
%\left(
\frac{X_{is}^* X_{ib} X_{js}^* X_{jb}}{(g^2 V_{ts}^* V_{tb})^2}
%\right)
\frac{M_W^2}{M_H^2}
\frac{f(m_i^2/M_H^2,m_j^2/M_H^2,M_W^2/M_H^2)}{S(m_t^2/M_W^2)},
\end{equation}
where $i,j = c,t$ and $f(x,y,z)$ and $S(x)$ are the Inami-Lim functions.
If we choose $Y_u^{\prime 33}$ appropriately,
the charged Higgs contribution to $M_{12}^s$ can vanish, %\footnote{
 keeping the contribution to $\Gamma_{12}^s$,
(In fact, if $Y_u^{\prime ij} \propto \delta_{ij}$ and $m_i \ll M_H$,
a GIM-like mechanism works for the dispersive part of the meson mixing amplitudes).
This is because the top-loop can contribute to $M_{12}^s$,
but not to $\Gamma_{12}^s$.
Therefore, one can avoid an excessive contribution to $M_{12}^s$
and still modify $\Gamma_{12}^s$ appropriately in this scenario.

\begin{figure}[t]
\center
\includegraphics[width=10cm]{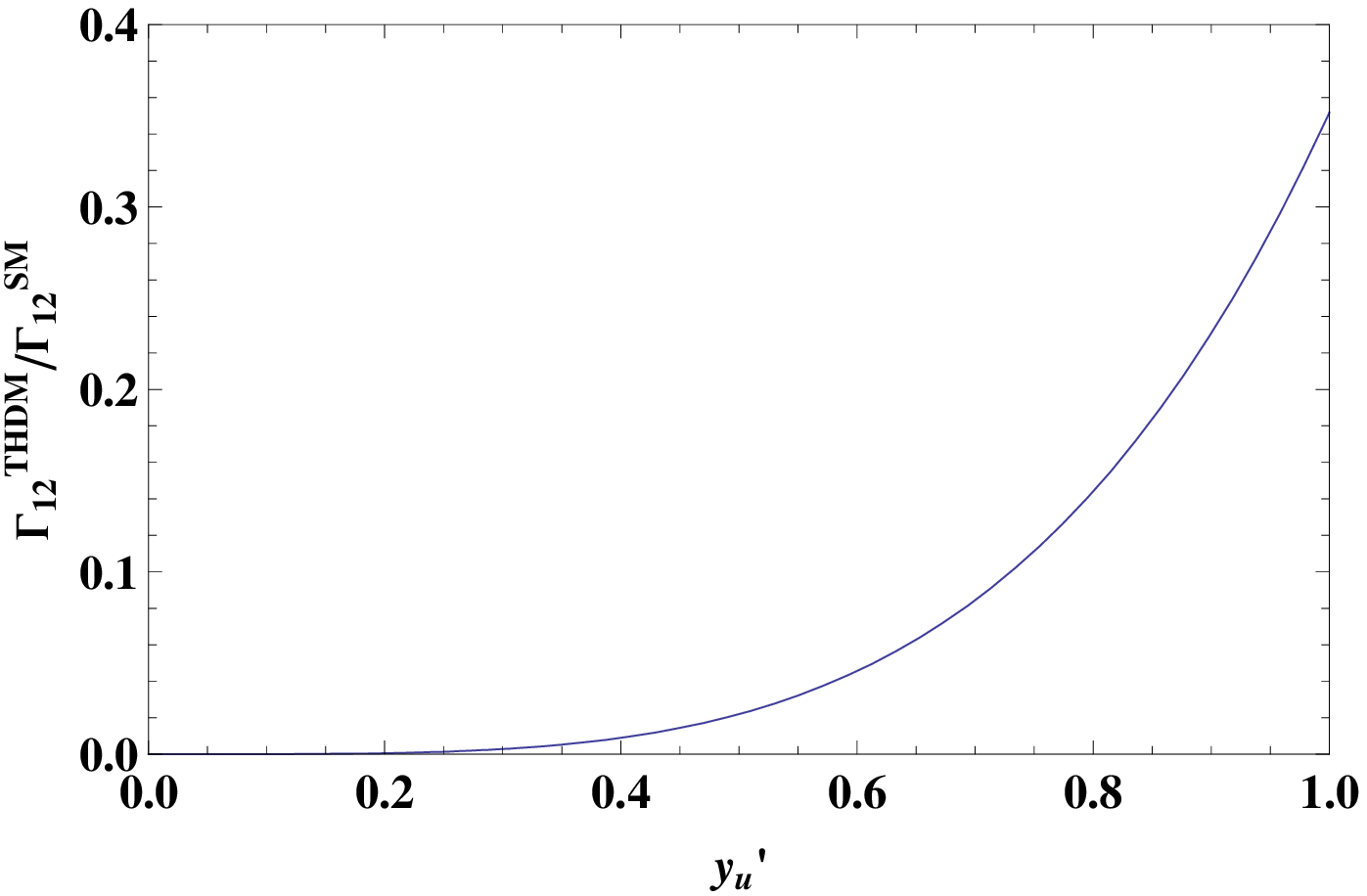}
\caption{$\Gamma_{12}^{\rm THDM}/\Gamma_{12}^{\rm SM}$
versus $y_u^\prime$
}
\label{Fig1}
\end{figure}

%(Show a graph of $\Gamma_{12}/\Gamma_{12}^{\rm SM}$)

In Fig.\ref{Fig1},
we show $\Gamma_{12}^{\rm THDM}/\Gamma_{12}^{\rm SM}$
as a function of $y_u^\prime$, where we assume $Y_u^{\prime ij} = y_u^\prime \delta^{ij}$
for simplicity.
We choose the charged Higgs mass to be 160 GeV.

In order to modify $\Gamma_{12}^s$, we need $Y_u^\prime$
whose magnitude
%The size of $Y_u^\prime$
depends on the charged Higgs mass $m_H$.
As previously mentioned, we need to fine-tune in order to obtain the proper up quark mass
in order to satisfy the lifetime ratio.
The fine-tuning can be relaxed when the charged Higgs is light.
However,
the constraints from $B^- \to \tau \bar\nu$ and
$B \to D \tau \bar\nu$ should be taken into
account \cite{Hou:1992sy}.
 In a general 2HDM, the non-holomorphic coupling $\ell \tau^c H_u^*$
can contribute to the $\tau$ mass, and thus the $\ell \tau^c H_d$ coupling
(which is important to $B \to \tau\bar\nu$)
may have freedom to relax the $B\to \tau\bar\nu$ constraint.

%Therefore,
%\begin{equation}
%\frac{M_{12}^{\rm NP}}{M_{12}^{\rm SM}} \sim \frac{M_H^2}{M_W^2}
%\frac{f(m_c^2/M_H^2)}{S(x_t)} \frac{\Gamma_{12}^{\rm
%NP}}{\Gamma_{12}^{\rm SM}}\frac1{\gamma_{cc}} \sim 0.6
%\frac{M_H^2}{M_W^2}\frac{\Gamma_{12}^{\rm NP}}{\Gamma_{12}^{\rm
%SM}}.
%\end{equation}
%
%This relation also restricts the charged Higgs mass. The heavy
%charged Higgs is disfavored, and the constraints from $B^- \to
%\tau \bar\nu$ and $B \to D \tau \bar\nu$ should be taken into
%account{ In general 2HDM, the non-holomorphic coupling $\ell
%\tau^c H_u^*$ can contribute to $\tau$ mass, and thus the $\ell
%\tau^c H_d$ coupling (which is important to $B \to \tau\bar\nu$)
%may have freedom to relax the $B\to \tau\bar\nu$ constraint. }.

The sizable Higgs coupling $Y_u^{\prime11}$
can provide an interesting hadron collider signal.
The coupling can cause a resonant production
of the charged and neutral Higgs bosons (which mainly contain $H_d$).
The Higgs bosons decay into two jets
via the $Y_u^{\prime 11}$ and $Y_u^{\prime 22}$ couplings,
which are needed to enlarge $\Gamma_{12}^s$.
Thus, dijet excesses can be observed
around the masses of the Higgs bosons.
If the Higgs boson is light ($\alt$ 200 GeV),
the dijet events are buried under the QCD background.
However, even if the Higgs boson is light, there is a chance to observe
dijet events produced by Higgs decays associated with $W$/$Z$
gauge bosons.
Recently, the CDF collaboration
has reported an excess of dijets associated
with $W$ boson (which decay into $\ell\bar\nu$) \cite{Aaltonen:2011mk}.
The excess in the dijet mass distribution is in the 120-160 GeV
range,
and it can be explained if the Higgs boson mass is
about 150-160 GeV (the peak of the dijet mass distribution
shifts to lower mass due to cuts).
We note that the bottom quark mass should be
generated by $q_3 b^c H_u^*$
because $b$-quark excess is not observed.

%there is a chance
%that the dijets produced by Higgs decays
%associated with $W$ gauge boson

\begin{figure}[t]
\center
\includegraphics[width=9cm]{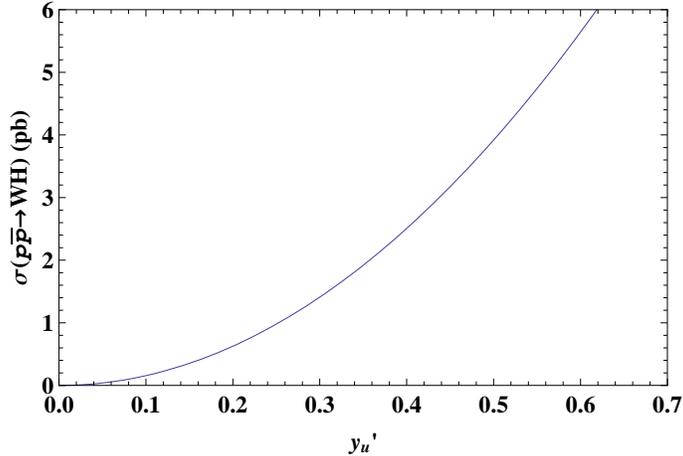}
\caption{$\sigma(p\bar{p}\rightarrow W H)$ plotted  as a function of $y_u^\prime$.
}
\label{Fig3}
\end{figure}

\begin{figure}[t]
\center
\includegraphics[width=10cm]{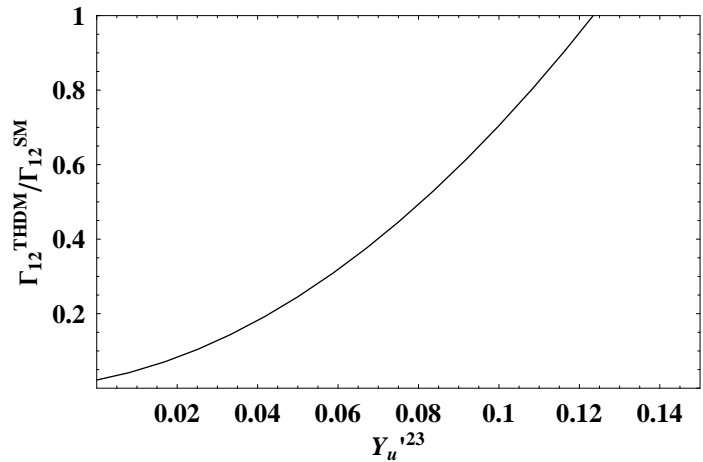}
\caption{$\Gamma_{12}^{\rm THDM}/\Gamma_{12}^{\rm SM}$
plotted as a function of $Y_u^{\prime 23}$ for $Y_u^{\prime 11} = Y_u^{\prime 22} = 0.5$.
}
\label{Fig2}
\end{figure}

%
%
%and the neutral Higgs can be produced directly via $u\bar u$, $d\bar d$ and $s\bar s$.
%
The charged and neutral Higgs bosons associated
with the $W$ boson
are produced by the $t$-channel exchange of the left-handed quarks
through the $Y_u^\prime$ coupling.
The resonant (or off-shell) production of the charged/neutral Higgs boson
through $Y_u^\prime$
can also contribute to the neutral/charged Higgs boson
associated with the $W$ boson.
Similar processes via a Higgs doublet
are also analyzed in \cite{Nelson:2011us,Cao:2011yt}.
%
%The resonant production of the charged and neutral Higgs bosons
%occur through the $Y_u^\prime$ coupling.
%
The CDF estimated production rate of $Wjj$ excess
is about 4 pb.
We use MadGraph/MadEvent (version 5) to estimate the $W H$ cross-sections in this model.
In Fig.\ref{Fig3}, we plot $\sigma(p\bar{p}\rightarrow WH)$ as a function of
$y^{\prime}_u$ for $m_{H^+} = 160$ GeV (neutral Higgs masses are also 160 GeV).
In the figure we use a $K$ factor $\sim 1.35$. The $WH$ production cross-section
is dominated by charged Higgs associated production (roughly 70\%).
The charged Higgs boson then decays 100\% into jets.
We find that  the excess can be explained for
Higgs couplings $Y_u^{\prime 11} \sim Y_u^{\prime 22} \sim 0.5$.
We also have $ZH$ (includes both charged and neutral Higgs) productions in this model,
but the cross-section is at least a factor 3 smaller compared to the $WH$ production.
The $Z H$ production cross-section goes further down if the neutral Higgs masses are
heavier than the charged Higgs masses.
%The ratio of $Wjj$ excess to $WW+WZ$ is about 0.25 in this model
%and found that the $W jj$ cross-section arising from $W^\pm H^\mp$ production
%is at least a factor of 3.5 bigger than the $Z jj$ productions
%(which involve $Z H^{\pm}$, $ZH^0$ and $Z A^0$) for Higgs masses ~160 GeV.
%The $Z jj$ production cross-section goes further down if the neutral
%Higgs masses are heavier than the charged Higgs masses.

In Fig.\ref{Fig2},
we plot
$\Gamma_{12}^{\rm THDM}/\Gamma_{12}^{\rm SM}$
as a function of $Y_u^{\prime 23}$,
where we choose
$Y_u^{\prime 11} = Y_u^{\prime 22} = 0.5$ and $m_{H^+} = 160$ GeV.
Since in this case %$Y_u^{\prime 11} \sim Y_u^{\prime 22} \sim 0.5$,
we need $Y_u^{\prime 23}$ to be $\sim 0.1$ to produce
the desired $\Gamma_{12}^s$ modification.
Due to this small $Y_u^{\prime 23}$, we have a very small
amount of single $b$ quark present in the $W jj$ signal.
As mentioned before, $Y_u^{\prime 23}$ is constrained by
$b_R \to s_L \gamma$ operator.
In order to turn on the $Y_u^{\prime 23}$ term,
one should make $Y_d^{33}$ small
and the bottom quark mass should come from the non-holomorphic term.
The absorptive part of the mixing is insensitive to $Y_u^{\prime 33}$,
and thus $Y_u^{\prime 33}$ can be used to adjust the
dispersive part of the mixings.

\section{Dimuon asymmetry in Left-Right Symmetric gauge theory}

In SUSY models,
the non-holomorphic terms ($Y_u^\prime, Y_d^\prime$ couplings) can arise only from
the finite corrections.
A sizable contribution to $M_{12}^s$ is easily obtained,
but in order to obtain sizable effects for $\Gamma_{12}^s$
consistence with other experimental constraints such as $b\to s\gamma$,
$O(100)$ TeV scale physics should be considered \cite{Dobrescu:2010rh}.

%In order to obtain the sizable effects to
%$\Gamma_{12}^s$

In the left-right symmetric gauge theory (LR model),
the gauge symmetry is extended to $SU(2)_L \times SU(2)_R \times U(1)_{B-L}$ \cite{moha}.
In this model,
a number of Higgs fields and couplings are usually needed in order to obtain
the required quark masses and CKM mixings.
The Higgs couplings can generate FCNC,
which can affect the dimuon asymmetry,
as described in the previous section.
In addition,
%In the LR model,
the $SU(2)_R$ gauge bosons can also contribute to
$\Gamma_{12}^s$ and $M_{12}^s$ \cite{Buras:2010pz}.
Let us estimate the
contributions of these gauge bosons to the mixing amplitudes.

%In this section, we obtain the possible amount of the $\Gamma_{12}^s$
%in LR model.

The RRRR operator $(\overline{q_R} \gamma_\mu b_R)(\overline{q_R} \gamma^\mu b_R)$
can contribute to $\Gamma_{12}^q$ ($q=d,s$):
\begin{equation}
\frac{\Gamma_{12}^{q \rm LR}}{\Gamma_{12}^{q\rm SM}}
\simeq
\left(\frac{V_{tq}^{R*}V_{tb}^R}{V_{tq}^{L*}V_{tb}^L}\right)^2
\left(\frac{g_R}{g_L}\right)^4
\left( \frac{M_{W_L}}{M_{W_R}} \right)^4 (1+O(m_c^2/m_b^2)),
\end{equation}
where
$V^L$ and $V^R$ are the left- and right-handed quark mixing matrices,
$g_{L,R}$ are the gauge coupling constants,
and $M_{W_{L,R}}$ are the masses of the $W_{L,R}$ gauge bosons
(in general, these bosons can mix).
We have used here the unitarity of the quark mixing matrices:
$V_{uq}^{*}V_{ub}+V_{cq}^{*}V_{cb}+V_{tq}^{*}V_{tb}=0$.
Note that the $M_{12}$ contribution from the RRRR operator is tiny
because of the unitarity of $V_R$ and $m_{u,c,t}/M_{W_R} \ll 1$.

The LLRR operator $(\overline{q_L} \gamma_\mu b_L)(\overline{q_R} \gamma^\mu b_R)$
can contribute to $M_{12}^q$:
\begin{equation}
\frac{M_{12}^{q \rm LR}}{M_{12}^{q \rm SM}}
\sim
\left(\frac{V_{tq}^{R*}V_{tb}^R}{V_{tq}^{L*}V_{tb}^L}\right)
\left(\frac{g_R}{g_L}\right)^2
\left( \frac{M_{W_L}}{M_{W_R}} \right)^2
\frac{2 A_2 (x_t^2,M_{W_L}^2/M_{W_R}^2)}{F(x_t^2)},
\end{equation}
where the loop functions $A_2$ and $F$ can be found in \cite{Nam:2002rq}.
Note that the contribution to $\Gamma_{12}^q$ via the LLRR operator
can be negligible
because charm or up quark masses are inserted
(or $W_L$-$W_R$ mixing is inserted twice).

As a result, we obtain
\begin{equation}
\frac{M_{12}^{q \rm LR}}{M_{12}^{q \rm SM}}
\sim
\sqrt{\frac{\Gamma_{12}^{q \rm LR}}{\Gamma_{12}^{q \rm SM}} }
\frac{2 A_2 (x_t^2,M_{W_L}^2/M_{W_R}^2)}{F(x_t^2)}.
\end{equation}
The correction to $\Gamma_{12}^q$
should be less than about 30\%
when $|M_{12}^{q \rm LR}/M_{12}^{q \rm SM}| \alt 2$
and $M_{W_L}^2/M_{W_R}^2 \alt 10^{-3}$.
Therefore,
the contribution via the right-handed gauge boson
is not a better choice to modify $\Gamma_{12}^s$
since modification to $M_{12}^s$ will be too large
irrespective of the choice of the right-handed quark mixing matrix.

The merit of the right-handed gauge boson is to modify
$\Gamma_{11}^d$ to adjust the lifetime ratio
for $B_d$ and $B_s$.
If the right-handed quark mixing matrix is
\begin{equation}
V^R = \left(
 \begin{array}{ccc}
   1 & 0 & 0 \\
   0 & 0 & 1 \\
   0 & 1 & 0
 \end{array}
\right),
\end{equation}
the operator $(\bar d_R \gamma_\mu b_R)(\bar c_R \gamma^\mu u_R)$
is generated by $W_R$ exchange,
and the lifetime of $B_d$ can be tuned depending on
$g^2_R/M_{W_R}^2$.
In the previous section,
we found that the element $Y_u^{\prime11}$ can
modify $\Gamma_{11}^d$ but it requires
fine-tuning to obtain the proper up quark mass.
The contribution from the right-handed gauge boson
can relax the fine-tuning.
We must adjust the contribution
to make the lifetime ratio lie within  the current experimental
uncertainty in eq.(\ref{lifetime-ratio}).
If we choose the elements (e.g. $V^R_{us}$) to be exactly zero, any
unwanted contribution to meson mixings from the $W_R$ gauge boson
can be avoided.

\section{Conclusion}

In this paper we have investigated the dimuon CP asymmetry
in $B_s$ decays in two Higgs doublet models.
We find  that the flavor changing neutral current (FCNC) induced
 by the Higgs couplings in  type III two Higgs doublet model can  enhance the decay width
$\Gamma_{12}^s$ and resolve the dimuon charge asymmetry puzzle,
consistent with all the  experimental constraints.
The enhancement of $\Gamma_{12}^s$ requires large Higgs couplings to the first
and second generations of quarks, which may help explain
the recent $3.2 \sigma$ $Wjj$ excess observed at CDF.

This work is partially supported by  DE-FG02-95ER40917 (B.D.),
STDF grant 437 and also ICTP Proj. 30 (S.K.), Excellent Research Projects of
National Taiwan University under grant number NTU-98R0526 (YM) and DE-FG02-91ER40626 (Q.S.).

\end{document}